\documentclass[prb,twocolumn,preprintnumbers,amsmath,amssymb]{revtex4}
\usepackage{graphicx}
\begin{document}

\title{Network equilibration and first-principles liquid water}

\author{M. V. Fern\'andez-Serra}
\email{mfer01@esc.cam.ac.uk}
\author{Emilio Artacho}
\affiliation{Department of Earth Sciences,
             University of Cambridge,
             Downing street, Cambridge CB2 3EQ, UK}
\date{9 July 2004}

\begin{abstract}
  Motivated by the very low diffusivity recently found in ab initio
simulations of liquid water, we have studied its dependence with
temperature, system size, and duration of the simulations.
  We use ab initio molecular dynamics (AIMD), following the
Born-Oppenheimer forces obtained from density-functional theory (DFT).
  The linear-scaling capability of our method allows the consideration
of larger system sizes (up to 128 molecules in this study),
even if the main emphasis of this work is in the time scale.
  We obtain diffusivities that are substantially lower than the
experimental values, in agreement with recent findings using similar
methods.
  A fairly good agreement with $D(T)$ experiments
is obtained if the simulation temperature is scaled down by $\approx$20\%.
  It is still an open question whether the deviation is due to the
limited accuracy of present density functionals or to quantum
fluctuations, but neither technical approximations (basis
set, localization for linear scaling) nor the system size (down to 32
molecules) deteriorate the DFT description in an appreciable way.
  We find that the need for long equilibration times is consequence 
of the slow process of rearranging the H-bond network (at least 20~ps 
at AIMD's room temperature).
  The diffusivity is observed to be very directly linked to
network imperfection.
  This link does not appear an artefact of the simulations, 
but a genuine property of liquid water.
\end{abstract}

% \pacs{XX.XX.xx}

\maketitle

\section{Introduction}

  After ten years of success of DFT-based AIMD simulations of
liquid water, including work on structural, dynamical, chemical and
electronic properties,\cite{Sprik96,Ortega96,Silvestrelli99jcp,
Silvestrelli99prl,parrinelloOH,parrinello-wet-e,Voth02}
recent papers\cite{Grossman04,Schwegler04,Asthagiri03} have questioned
some of the results of those early studies, showing that if the
simulations are allowed to run longer, the diffusivity drops by one
order of magnitude and the liquid becomes over-structured.
  The origin of the discrepancy with 
experiments\cite{waterexpD,waterexp1,waterexp2} is still unclear.

  There are obvious limitations in the AIMD description of 
liquid water that could account for them, including the inability of 
present gradient-corrected (GGA) density functionals to describe dispersion 
interactions, or the complete neglect of quantum fluctuations
in the classical treatment of nuclear dynamics.
  The former problem has hardly been addressed and demands simulations
where van der Waals interactions are explicitly accounted for.
  The recent DFT proposals that include these 
interactions\cite{Kohn98,Lundqvist04} are still too demanding to 
allow realistic AIMD simulations of this sort.
  Empirical force fields have an enormous advantage here, since those
interactions can be reasonably well described with little effort.
  For the latter problem, the complete quantum mechanical treatment of
both electronic and ionic degrees of freedom is still computationally
too costly, and, even though some pioneering studies have recently
appeared,\cite{parrinelloqtm,kleinqtm} their approximations have to
be pushed to the limit and their reliability is still 
unclear.\cite{Schwegler04}
  Empirical simulations including proton quantum effects are again 
much more feasible.

  Even if a wide range of empirical potentials exist for pure liquid water
that offer a better description than that attainable nowadays by DFT,
it is extremely important to have a working description of
liquid water at the DFT level, not for the study of water itself, but
for that of systems interacting with water.
  This is important in scientific fields as wide as wet chemistry,
biochemistry, geochemistry and environmental sciences. 
  Empirical potentials do not have enough flexibility and transferability 
to describe the large variety of processes happening in wet systems.

  The purpose of this work is to assess the situation regarding 
the equilibrium description of DFT water, as well as understand the 
equilibration process that lurks behind the problems in reaching it.
  We present in the following results of simulations for different 
sizes, at different temperatures, and for relatively long times.
  Because the long-term aim is using DFT water in interaction with
other systems, the scalability of the DFT description becomes crucial.
  We have thus used a method based on numerical atomic orbitals 
of finite support that allows linear-scaling DFT 
calculations\cite{SiestaPRBRC,SiestaJPCM} and therefore the 
possibility of much more efficient treatment of larger system sizes.
  The method is validated below for DFT liquid water, including
the localization approximations required for linear
scaling.\cite{SiestaJPCM,Ordejon95,Kim-mauri-galli}
  After the characterization of the equilibrium properties, we present 
results on non-equilibrium relaxation processes, which provide insights 
into why the simulations need longer times, how to look at the DFT
deficiencies, and, more importantly, into the nature of liquid
water itself.

\section{Method}

% --------- General

  Our simulations are performed using the self-consistent
Kohn-Sham approach\cite{Kohn-Sham} to density-functional 
theory\cite{Hohenberg-K} in the generalized-gradient approximation (GGA).
  The BLYP\cite{blyp1,blyp2} exchange-correlation functional was chosen
following previous studies,\cite{Sprik96} even if the reported performance 
for liquid water seems to be quite similar among GGA 
functionals\cite{Grossman04} (a more detailed comparison using 
different functionals will be presented elsewhere).

  Core electrons are replaced by BLYP-generated norm-conserving 
pseudopotentials\cite{Troullier-Martins,inp} in their fully non-local 
representation.\cite{Kleinman-Bylander}
  Numerical atomic orbitals (NAO) of finite support are used as basis set,
and the calculation of the self-consistent Hamiltonian and overlap matrices 
is done using the linear-scaling {\sc Siesta} 
method.\cite{SiestaPRBRC,SiestaJPCM}
  Integrals beyond two-body are performed in a discretized real-space
grid, its fineness determined by an energy cutoff of 150 Ry .

  The solution of the eigenvalue problem is performed either with
the linear-scaling solver of Kim {\it et al.}\cite{Kim-mauri-galli} or by 
diagonalization.
  The former is more efficient for larger sizes (due to the cube scaling
of the latter), but imposes an additional localization approximation
on the basis functions for the occupied Hilbert 
space.\cite{Ordejon95,Kim-mauri-galli}
  The effect of this approximation in our system is assessed below.

% --------- Basis General

  The NAO bases were variationally optimized for the water 
molecule.\cite{javibases,edubases}
  The double-$\zeta$ polarized (DZP) level was found to offer a
good balance between accuracy and efficiency.
  For this system it means two 2$s$ orbitals, two 2$p$ shells and 
one 3$d$ shell for oxygen, and two 1$s$ orbitals and one 2$p$ shell 
for hydrogen.
  Three basis sets were tried at this level, differing in the 
cutoff radii of
the support regions of their wave-functions.
  These radii are controlled by a single ``pressure" 
parameter,\cite{edubases} by which tighter orbitals 
are obtained with higher pressure.
  Three pressures were considered (0.0 GPa, 0.2 GPa, and 0.5 GPa), and
their performance in the description of the water monomer and dimer
is shown in Table~\ref{dimer}. 

\begin{table}
\caption[]{Selected properties of the water monomer and dimer calculated
with our method for three different basis sets. They are compared to 
results of plane-wave (PW, 70 Ry energy cutoff) and well converged 
Gaussian (GTO) calculations, and to experiment. The BLYP functional was 
used in all the calculations. $d$ stands for dipole moment, $E_b$ for 
binding energy and BSSE for basis-set superposition error.}
\begin{ruledtabular}
\begin{tabular}{lcccccccc}
 &\multicolumn{3}{c}{Monomer}& & \multicolumn{4}{c}{Dimer}\\
Basis& $r_{\rm OH}$(\AA) & $\widehat{\rm HOH}$ & $d$(D) & &
$r_{\rm OO}$(\AA) & $\widehat{\rm OHO}$ & $E_b$(eV) & BSSE \\ 
\hline 
DZP$^{0.5}$ &0.967   &104.4   &2.13   & &2.92   &175.1   &4.76&$12\%$ \\
DZP$^{0.2}$ &0.970   &104.2   &2.04   & &2.94   &177.1   &4.68&$20\%$ \\
DZP$^{0.0}$ &0.974   &104.0   &1.89   & &2.95   &178.1   &4.01&$30\%$ \\
PW$^a$      &0.973   &104.4   &1.81   & &2.95   &173.0   &4.30 & - \\ 
GTO$^b$     &0.972   &104.5   &1.80   & &2.95   &171.6   &4.18 & - \\
Exp.&0.957$^c$ &104.5$^c$   & 1.85$^d$ & &2.98$^e$ & 174.0$^e$ &5.44$^f$& - \\
\end{tabular}
\footnotetext{
$^a$Ref.~\onlinecite{Sprik96};
$^b$Ref.~\onlinecite{Schwegler04};
$^c$Ref.~\onlinecite{wbenedict75};
$^d$Ref.~\onlinecite{sclough73};
$^e$Ref.~\onlinecite{jodutola80};
$^f$Ref.~\onlinecite{lcurtiss79}.}
\end{ruledtabular}
\label{dimer}             
\end{table}

% --------- Basis choice & dipole

  For the AIMD simulations we opted for the intermediate basis set (0.2 GPa),
even if some numbers in the table are better reproduced by the long 
one (0.0 GPa).
  On one hand, the efficiency of the calculations rapidly degrades with 
longer basis-function cutoff radii.
  On the other, the importance of long basis-function tails for the monomer 
and dimer stems from their gas-phase character.
  In the liquid phase these long tails become irrelevant given the 
presence of basis functions in neighboring molecules.
  This effect is quite apparent in the behavior of the dipole moment.
  Table~\ref{dimer} shows a clear tendency of growing overestimation
of the dipole moment with tighter basis functions.
  However, the 2.04 D value obtained for the intermediate basis
decreases to 1.74 D if calculated with the 
same basis set, but surrounded by the basis functions of four neighboring 
(absent) molecules.
  The long basis has the additional disadvantage of a large 
basis set superposition error (BSSE) in the hydrogen-bond (HB) description,
while the short basis already shows appreciable discrepancies that are
unlikely to be corrected by the presence of NAOs in neighboring molecules.

% --------- g(r)

\begin{figure}[t!]
\includegraphics*[scale=1.5]{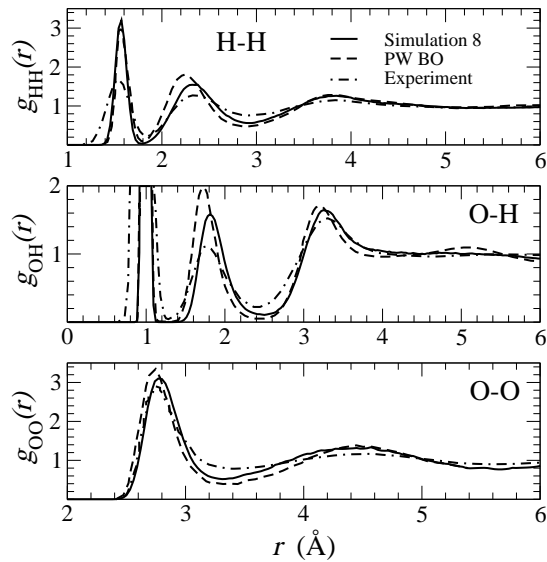}
\caption[]{Comparison of the H--H, O--H, and O--O radial distribution 
functions as obtained in this work for 64 molecules (solid line), with plane 
waves\cite{Schwegler04} (dashed line, PW BO stands for plane waves 
using Born-Oppenheimer forces), and by experiment\cite{waterexp1,waterexp2} 
(dot-dashed line), at a temperature of 300~K.}
\label{grcomp}
\end{figure}

  In Fig.~\ref{grcomp}, the results for radial distribution functions (RDFs)
for liquid water as produced by our method, are compared with those of 
experiment\cite{waterexp1,waterexp2} and those of recent AIMD
 Born-Oppenheimer 
results,\cite{Schwegler04} using a very similar method based on BLYP, 
norm-conserving pseudopotentials and a plane-wave (PW) basis. 
  There are clear discrepancies between the two theoretical methods,
our results showing an overall less structured liquid.
  Our slightly longer O-O distance along the HB seems
to correlate with a shorter intramolecular O-H distance (a weaker
HB is expected for a stronger intramolecular bond).

  The differences have to be ascribed to incomplete basis sets,
certainly in the NAO side, but also quite likely in the PW side.
  PW cutoffs in the range of 90 Ry or lower, as used in many PW studies 
(a 85 Ry cutoff was used in Ref.~\onlinecite{Schwegler04}),
are not extremely converged for GGA norm-conserving oxygen 
pseudopotentials, but rather represent a sensible 
compromise between accuracy and efficiency.
  Most important, however, is the fact that both simulations deviate from
experiment in a very similar way, both displaying a clearly over-structured
liquid.
  This over-structuring trend correlates with the very low diffusivities
found by both methods, as discussed below.

% -------- Linear Scaling

\begin{figure}[t!]
\includegraphics*[scale=1.5]{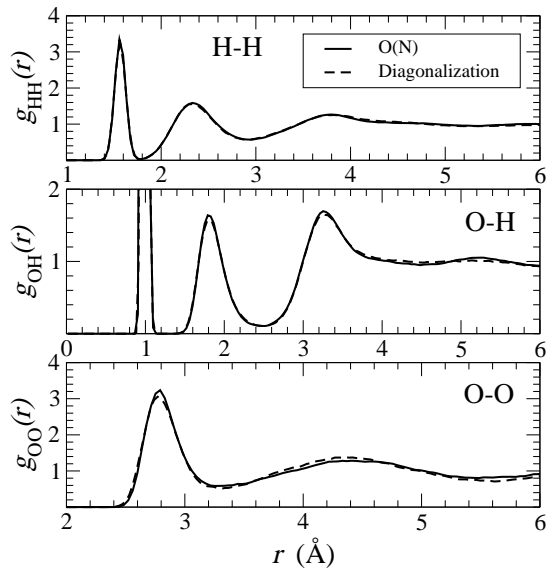}
\caption[]{Comparison of the RDFs obtained with diagonalization (solid 
line) and the $O(N)$ solver of Kim {\it et al.}\cite{Kim-mauri-galli} 
(dashed line) at a temperature of 300 K.}
\label{ON}
\end{figure}

  The approximations above allow the linear-scaling computation of the
Hamiltonian and overlap matrices.
  The $O(N)$ equivalent of the diagonalization demands an additional
localization approximation.
  The $O(N)$ solver of Kim {\it et al.}\cite{Kim-mauri-galli} does it
by imposing finite support to the Wannier-like localized solution 
wave-functions it finds.

  Figure~\ref{ON} tests the effect of the extra localization approximation
by comparing the RDFs obtained using diagonalization and using the $O(N)$ 
solver in a 64-molecule 15-ps simulation. 
  The confinement radius for the localized solution wave-functions is
5.0 \AA.
The $O(N)$ simulation corresponds to simulation 6 in Table~\ref{table2} 
and the diagonalization one to simulation 8.
 The comparison is very satisfactory, the differences being substantially 
smaller than when comparing with experiment.
  From a practical point of view, the point at which the linear-scaling
solver begins to be of advantage computationally, for this system and
this basis, is around 32 molecules.
  Since the main emphasis here is on long time scales, the simulations
presented below are up to 128 molecules only. 
  For these sizes, diagonalization is still affordable and is the method
chosen for the simulations below, in our aim to provide here as clean-cut
results as possible.
  It is, however, an important consequence of this study, the new perspectives
opened by the possibility of using linear-scaling for larger wet systems.
  
% -------- Simulation details

\begin{table}
\caption[]{AIMD simulations performed in this work. $N$ stands for the 
number of molecules, $a$ for the box size, $T$ for final 
equilibrated temperature, $\tau_{sim}$ for the AIMD simulation time after 
AIMD equilibration, $\tau_{eq}$ for the AIMD equilibration time,
``Model" for the model used for preparation, $T_{pre}$ for the
temperature at which the preparation model had been equilibrated and
$T_i$ for the AIMD initial temperature (after the $\tau_{eq}$ anneal).
Temperatures in K and times in ps.}
\begin{ruledtabular}
\begin{tabular}{crrccclcc}
\# & $N$ & $a$ (\AA) & $T$ & $\tau_{sim}$ & $\tau_{eq}$ & 
Model & $T_{pre}$ &$T_i$ \\ 
\hline
1  & 32  & 9.865  & 298 & 20 & 4 & AIMD  & 315 &300\\
2  & 32  & 9.865  & 315 & 32 & 6 & SPC/E & 300 &300\\
3  & 32  & 9.865  & 325 & 20 & 4 & AIMD  & 300 &335\\
4  & 32  & 9.865  & 345 & 30 & 4 & TIP5P & 325 &325\\
5  & 32  & 9.890  & 305 & 20 & 4 & AIMD  & 315 &300\\
6  & 64  & 12.417 & 297 & 15 & 4 & AIMD  & 320 &300\\
7  & 64  & 12.417 & 326 & 24 & 4 & SPC/E & 315 &300\\
8  & 64  & 12.460 & 303 & 20 & 4 & AIMD  & 320 &300\\
9  & 128 & 15.710 & 320 & 11 & 3 & SPC/E & 300 &300\\
\end{tabular}
\end{ruledtabular}
\label{table2}             
\end{table}

  We have carried out a series of nine MD simulations of heavy water
with varying size, density, temperature, and equilibration process.
  Their details are given in Table~\ref{table2}.
  AIMD equilibration is accomplished by means of temperature annealing 
(velocity re-scaling),\cite{allen-tildesley} while the actual simulations are 
performed by straight Verlet's integration,\cite{allen-tildesley} given our
interest in dynamical quantities.
  In all the simulations (both empirical, see below, and ab initio) the 
time step used was 0.5 fs.
  The observed total-energy drifts corresponded to drifts in the system 
temperature between 0.26 K/ps and 0.36 K/ps.
  The different (final) temperatures in Table~\ref{table2} are the result of 
different relaxation processes, not only in the preparation and 
further AIMD equilibration, but, most importantly,
during $\tau_{sim}$ itself.
  Instead of the usual approach of long enough equilibration times
and only monitoring the trajectories once well equilibrated, we chose
to explore the long time-scale equilibration process itself, by 
monitoring the non-equilibrium part of the simulation, as discussed 
below.

  The simulations were performed at constant volume (fixed cell size and
shape, under periodic boundary conditions). 
  The slight dispersion in the system densities considered in the literature
suggested the study of the effect of slight density changes (below 1\%) in 
the dynamical and structural results for our DFT water.
  Consistently with expectations (see Ref.~\onlinecite{Vaisman93}
for the density dependence of the diffusivity), we found such effects of 
marginal importance for the present study and will not be further 
discussed here.

  Empirical simulations were performed using different force fields
(TIP5P\cite{tip5p} and SPC/E\cite{SPC/E}) in order to prepare reasonably 
equilibrated starting points for AIMD.
  All these simulations were performed with the GROMACS MD 
package\cite{gromacs1,gromacs2} under constant volume and temperature 
conditions using a Berendsen-type thermostat.\cite{berendsen}
  We equilibrated the simulations during 200 ps for 32 and 64 
molecules and 150 ps for 128 molecules.

\section{Results and Discussion}

\subsection{Temperature dependence}

\begin{figure}[t!]
\includegraphics*[scale=1.5]{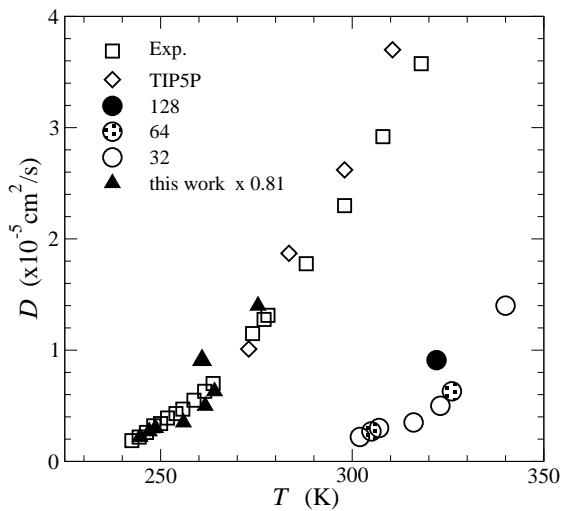}
\caption[]{Diffusivity vs temperature for this work
(circles), experiment\cite{waterexpD} (squares), TIP5P\cite{tip5pDif} 
(diamonds) and our AIMD data with a temperature re-scaling of 19\% 
(triangles).}
\label{diffT}
\end{figure}

  In Figure~\ref{diffT} the temperature dependence of our computed 
diffusivity is presented and compared with experimental values
at similar and lower temperatures\cite{waterexpD} and the corresponding 
values for the TIP5P potential.\cite{tip5pDif}
  The diffusion coefficient is calculated using the Einstein relation:
\begin{equation}
6D=\lim_{t\rightarrow\infty}\frac{d}{dt}
\langle\left|r_i(t)-r_i(0)\right|^2\rangle.
\end{equation}
  Eq.~(1) is evaluated computing the mean square displacement (MSD) of the 
oxygen atoms for multiple initial configurations equally spaced by 5 fs.

  Confirming previous results,\cite{Grossman04,Schwegler04} the figure 
displays an underestimation of the room-temperature diffusivity of around 
one order of magnitude.
  It can also be seen as an overestimation of the temperature needed
in the simulation to reach a certain diffusivity.
  Indeed, the same AIMD diffusivities plotted against a 19\%-scaled-down
simulation temperature give a quite acceptable agreement with experiment
(triangles in the figure).
  The implication is that our AIMD description overestimates by
around 25\% the features of the energy landscape 
relevant for diffusion and flow.

\begin{figure}[t!]
\includegraphics*[scale=1.5]{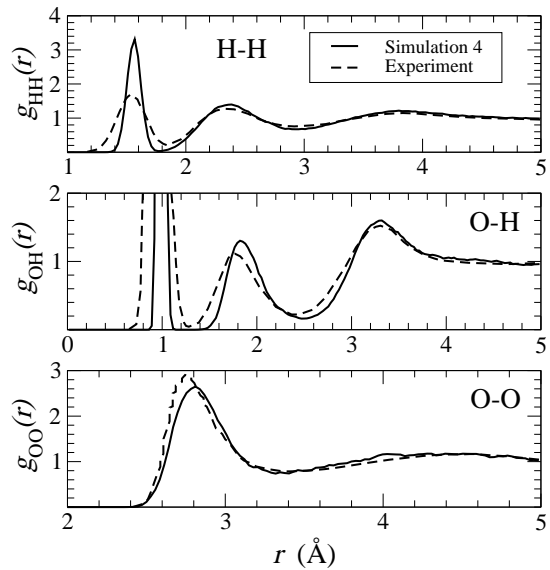}
\caption[]{Comparison of the H--H, O--H, and O--O RDFs 
as obtained in this work at 345~K (solid line), and by 
experiment\cite{waterexp1,waterexp2} at 300~K (dashed line).}
\label{exp-350}
\end{figure}

  One could then perform AIMD simulations at a higher temperature
in order to describe a liquid with a diffusivity comparable to 
room-temperature experiments.
  In Fig.~\ref{exp-350} we compare the room-temperature experimental
RDFs with our corresponding results for a temperature of 345 K (the highest 
considered here), and the improvement is evident.
  It is unclear, however, what kind of agreement one would obtain for
other properties, since the local dynamics is controlled by
atoms moving at velocities corresponding to the actual temperature
of the simulations.

  Schwegler et al.\cite{Schwegler04} reported a temperature overestimation
that required scaling down by 28\% (25\% if the simulations use the 
Car-Parrinello scheme), slightly larger than our 19\%, but clearly 
displaying the same trend.
  The higher diffusivity obtained with our NAOs as compared
with PW, correlates with the less structured RDF for the NAOs in 
Fig.~\ref{grcomp}.
  Both discrepancies point to a weaker HB in the liquid 
for our description, as discussed in the previous section.

  The most important result, however, is that both NAOs and PW disagree
with experiment more substantially than among themselves.
  This shows that the main problem with the DFT description of liquid water
is not in the technical approximations used in either method, but
in the more fundamental approximations, namely, the GGAs (BLYP in this
case) and/or the neglect of quantum fluctuations.
  
  We experimented with different flavors of GGA, and in general we agree
with previous reports in the conclusion that results for PBE\cite{PBE} 
do not change the main findings for BLYP.
  It is clear that semi-local exchange-correlation functionals like these 
ones miss the non-local correlation effects that give rise to 
dispersion forces. 
  However, since the origin of the discrepancy has not been ascribed 
to non-local effects, we decided to try other flavors, in the spirit of 
finding an efficient approach that works.
  A detailed study will be presented elsewhere, but preliminary results 
for the variant version of PBE proposed by Hammer {\it et al.}\cite{rpbe}, 
called RPBE, showed promisingly higher diffusivities ($\approx 1.4 
\times 10^{-5}$ cm$^2$/s at 300~K), with RDFs very similar to experiment
if not under-structured.

\subsection{Size effects}

  System-size effects have been addressed before\cite{Grossman04,waterexp2} 
using mainly empirical potentials, due to the difficulty of studying 
larger sizes with AIMD. 
  These studies indicated that size effects seemed not to be the
problem, but the need for confirmation of this conclusion using AIMD 
was pointed out.\cite{Grossman04,waterexp2} 
  Not only because of the difference between empirical and ab-initio 
forces, but also because empirical MD approaches tend to impose radial 
cutoffs to the attractive $R^{-6}$ potentials in order to avoid that an
atom interacts with its periodic images.
  Simulations for the 32-molecule system impose quite a substantial
and abrupt cut in the interactions, which originates a specific size effect,
irrelevant to our problem, and possibly masking the interesting effects.

\begin{figure}[t!]
\includegraphics*[scale=1.5]{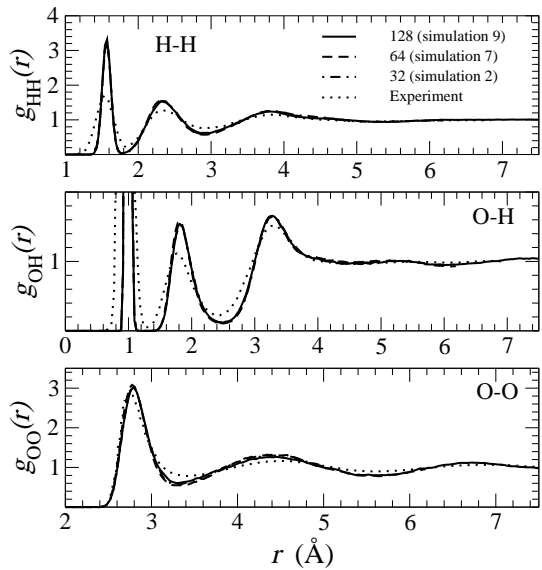}
\caption[]{Comparison of the H--H, O--H, and O--O RDFs obtained 
using 128 molecules (solid line), 64 (dashed line), 32 (dot-dashed line) and 
experiment\cite{waterexp1,waterexp2} (dotted line).}
\label{size}
\end{figure}

  We have thus carried out simulations with 32, 64 and 128 water molecules.
  A comparison of their RDFs is shown in Fig.~\ref{size}, where we do not 
find significant differences between the three sizes, supporting the 
conclusions of previous studies.\cite{Grossman04}  
  Furthermore, even if the 64-molecule and the 128-molecule simulations give 
slightly higher diffusivities than the 32-molecule one, it is apparent in 
Fig.~\ref{diffT} that the size effect produces marginal errors in the 
diffusivity as compared with the more substantial ones discussed before.

  This absence of more substantial size effects could be taken as  
indication that the structure of water is mainly determined by short 
range forces.\cite{Grossman04}
  The validity of such conclusion depends on what is understood by
``structure".
  We have to keep in mind that in these simulations the density is 
fixed to the experimental value.
  An AIMD simulation of variable cell size (a large system size would be
required to reduce the pressure and volume fluctuations) would give
a theoretical density that could appreciably differ from the experimental 
one, not least because of the neglect of dispersion forces.
  Indeed, we do observe in our simulations an average positive value 
of 2-4 kBar for the simulated pressure. 
  Changes in the density will not show anything of substance in the
nearest-neighbor peaks of the RDFs, since they are mainly determined by 
the HBs, but they will affect farther structural features.

\begin{figure}[t!]
\includegraphics*[scale=1.5]{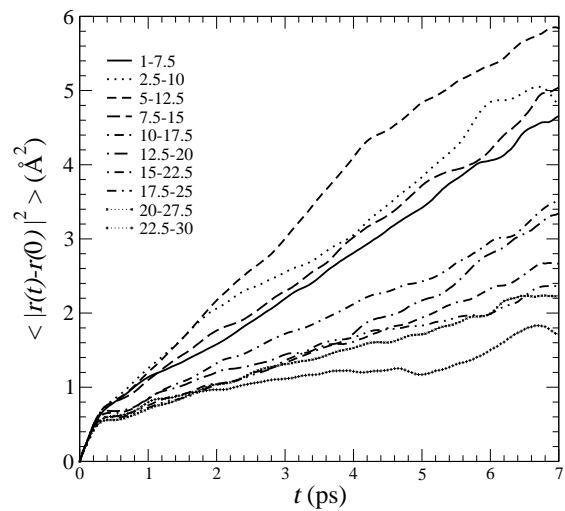}
\caption[]{Mean square deviation [as $F(t)$ in Eq.~(2)] for the oxygen 
atoms vs time for simulation 2. The plots are computed in
7.5-ps-wide windows every 2.5 ps.}
\label{diff-window}
\end{figure}

\subsection{Equilibration}

  It took ten years to find out about the problem of DFT 
water discussed above. 
  That fact in itself points to a different problem, or rather a 
combination of two, both addressed in this section: $(i)$ there is a 
long time scale associated to specific relaxation processes, and $(ii)$
it can be difficult to observe them.
  Starting by the latter, it is customary to obtain the diffusivity $D$ 
as in Eq.~(1), from the slope of the function $F(t)$ defined in the form of 
an integral of the MSD instead of the MSD itself,
\begin{equation}
F(t)=\int_{t_r=0}^{t_{max}}\langle|r_i(t-t_r)-r_i(t_r)|^2\rangle
\end{equation}
over a time window ($t_{max}$), normally the total of the simulation time.
  This procedure produces a substantial reduction of the fluctuations
allowing a better determination of the required slope.
  If an equilibration process is still taking place and the 
diffusivity is intrinsically not stabilized yet, instead of a 
time evolution of that slope (a bend), still a (roughly) linear curve 
is obtained, with an average slope. 
  It misleads to a reading of such slope as that of an equilibrated system.
  
  Instead, we have chosen to consider our simulations in a sequence of
overlapping time windows, each of them long enough (7.5 ps), to ensure
enough range for the linear behavior to be extracted from the 
MSD averaged plot.
  The $F(t)$ curves are shown in Fig.~\ref{diff-window}.
  Obtaining a slope from them is still tricky and not devoid of 
ambiguities, including the choice of start and end times within 
the time window for the linear fit.
  We have checked, however, that, if systematic and carefully done, 
the variability in the extracted values does not affect the results 
in a substantial way for the purposes of this study.
  This ``moving-window" approach allows observing the evolution of the
diffusivity with time, and address the equilibration problem.

\begin{figure}[t!]
\includegraphics*[scale=1.5]{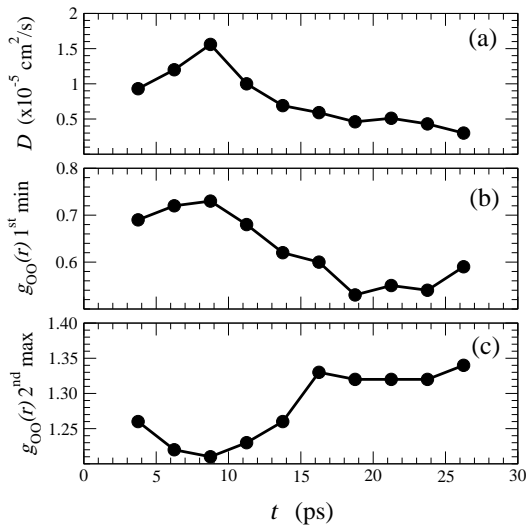}
\caption[]{Evolution of the diffusivity (a), the height of the first minimum
(b), and of the second maximum for $g_{\rm OO}(r)$ (c), in simulation 2.
Calculated in 7.5-ps-windows every 2.5 ps.}
\label{evolution}
\end{figure}

  In Fig.~\ref{evolution} we show the evolution of the diffusivity 
and relate it to the evolution of the liquid structure, as monitored by
he heights of the first minimum and of the second maximum of 
$g_{\rm OO}(r)$.
  The figure shows a clear non-equilibrium behavior, both in the 
diffusivity and in the structural characteristics, in a time
scale of tens of picoseconds, of the order 20-30 ps.
  Longer simulations would be needed to be more precise.
  We do not think it is justified at this point to determine
the equilibration time more precisely, bearing in mind 
that it varies with temperature and possibly with size as well.
  A consequence of this result that should be kept in mind is that
the ``equilibrium" properties obtained in the previous section,
as well as what obtained in previous studies (to our knowledge, 
the longest AIMD runs reported are not longer than 20-30 ps), 
are not necessarily completely equilibrated (as seen in 
Fig.~\ref{evolution}), and should be taken with caution.
  History dependence is thus to be expected in similar AIMD 
simulations, {\it i.e.}, dependence on the preparation model and 
initial temperature.

  It is tempting at this stage to relate our equilibration time 
with the one observed by inelastic UV scattering\cite{Sette04} for
the structural relaxation probed by sound modes in the liquid,
which, from values lower than 1 ps at room temperature, increases
to higher than 20 ps below 250~K.
  Considering the 20\% down-scaling discussed above, our 
room-temperature AIMD relaxation time scale would be consistent with the 
experimentally measured characteristic time for $\approx$240~K.
  If that is the case, accurate AIMD simulations (with no need for
temperature re-scaling) should equilibrate in $\approx$1 ps.
  The same would be true for our AIMD at $T\approx$375~K.

\begin{figure}[t!]
\includegraphics*[scale=1.2]{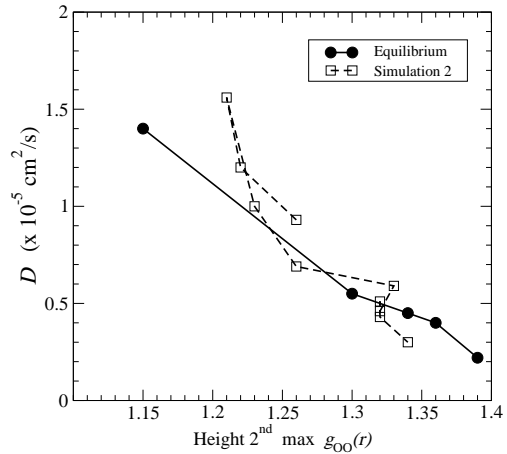}
\caption[]{Diffusivity versus height of the second maximum in 
$g_{\rm OO}(r)$. Filled circles: equilibrium results for all the 
simulations with 32 molecules. Open squares: non-equilibrium
evolution of the simulation 2 illustrated in Fig.~\ref{evolution}.}
\label{DvsS}
\end{figure}

  Fig.~\ref{evolution} also shows a clear correlation between 
diffusivity and structure, very similar to what observed in equilibrium.
  This correlation is made explicit in Fig.~\ref{DvsS}, where 
the non-equilibrium behavior for the relaxation shown in 
Fig.~\ref{evolution} is compared with the equilibrium plot
of $D$ versus the height of the second maximum of $g_{\rm OO}(r)$,
as obtained in all the equilibrated simulations with 32 molecules.
  
  This graph has interesting implications.
  The long equilibration time scale seems to be related to finding
the equilibrium for some structural characteristic, $X$, 
but the diffusivity seems to equilibrate on a much shorter time scale 
to the instantaneous state of such $X$.
  This situation could  be interpreted in terms of two time scales, 
one long, in which $X$ evolves toward equilibrium, and a shorter 
one within which everything else happens, including diffusion. 

  Of course, this is an idealization, as is most clearly seen
in the third point of Fig.~\ref{evolution}(a), that corresponds to the
highest open square in Fig.~\ref{DvsS}.
  This and the other smaller deviations from the equilibrium curve in
Fig.~\ref{DvsS} indicate that the equilibration of $D$ to the corresponding 
value for $X(t)$ is not strictly instantaneous.
  This simple interpretation offers fruitful insights, however, which 
are explored in the following, by analysing the liquid structure in
terms of a HB network.\cite{stillingerwater}

\subsection{Hydrogen-bond network}

  The molecules in water bind to each other by hydrogen bonds (HBs).
  Even if the character of this kind of bond is still 
controversial,\cite{Guo02,Romero01,Ghanty00,hbondpaper} two important 
features are clear, namely, its strength (between that of a covalent bond 
and a Van der Waals one\cite{stillingerwater}), and its directionality. 
  The chemical tendency is for each molecule to be surrounded by
four others, donating two HBs and receiving two, in a tetrahedral arrangement.
  This tendency is perfectly satisfied in the crystalline phases of
ice, but the HB network in liquid water at any given time
is imperfect, with many four-coordinated molecules (even if the
HBs may be stretched or bent), but some under-coordinated
and over-coordinated ones as well.
  The picture is dynamic, with a continuous breaking and forming of
HBs, with an average bond life-time of the order of 1 ps.\cite{Luzar00}
  In this work we describe the HB network mainly by its coordination
defects, namely, the under- and over-coordinated molecules.

\begin{figure}[t!]
\includegraphics*[scale=1.5]{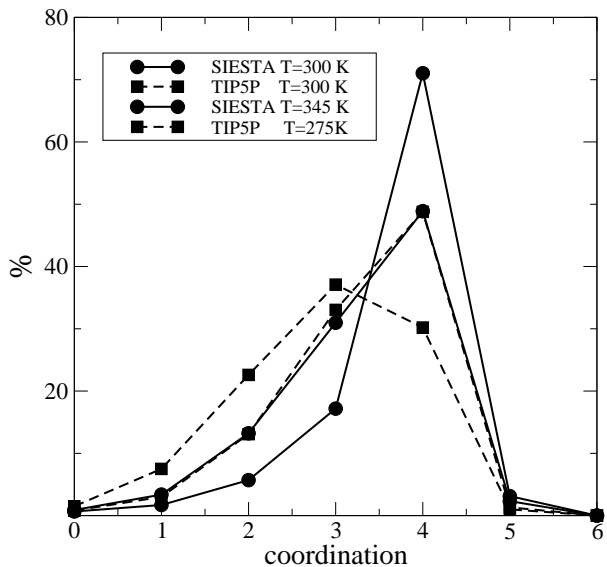}
\caption[]{Distribution of molecules with different coordinations.
AIMD results (circles) are compared with results for the TIP5P 
potential\cite{tip5p} (squares). Comparison for the same temperature (300~K 
for both) is presented, as well as for 20\% re-scaled temperatures (275~K
for TIP5P and 345~K for AIMD; these are the ones practically superimposed).}
\label{coord-hist}
\end{figure}
   
  In order to characterize the HB network a criterion must
be adopted to decide whether two molecules are bonded or not.
  The usual criterion relies on two aspects, $(i)$ an oxygen-oxygen 
distance smaller than some cut-off value, normally\cite{Chandler96,Starr00} 
the minimum after the first peak in $g_{\rm OO}$(r); and $(ii)$ 
a HB angle larger than an arbitrary minimum value.
  Following Refs.~\onlinecite{Chandler96,Starr00} we adopted a minimum
angle of 145$^{\rm o}$.
  For the characterization of network defects, we have also resorted to
a temporal criterion for the definition of a HB, since fluctuations
of distances or angles close to the critical values would otherwise
appear as short-lived coordination defects, masking the defect statistics 
we want to monitor.
  Keeping track of HBs with life-times longer than typical vibrational
or librational periods is enough for the purpose (we used a 
threshold of 250 fs, as defined by the cage effect\cite{Gallo96}).

  Fig.~\ref{coord-hist} shows the equilibrium distribution of molecules with
different coordinations.
  Since there are no direct experimental results to compare to, 
comparison is presented with the results of the TIP5P force field.
  Note the asymmetry of the distribution, everything
happening between the ideal coordination and under-coordination.
  Notice, however, that imposing a threshold life time in our
HB definition biases the distribution toward lower coordinations
(a longer minimum life time implies less HB qualifying as such, 
and thus the molecules are less coordinated).
  Nevertheless, we have checked the variation of the distributions 
with different life-time thresholds, finding that the overall shape is 
robust, maintaining the observed asymmetry.
  First, same temperatures are compared, both AIMD and TIP5P
at 300~K, showing a very important difference, with AIMD displaying
less than 30\% defects while TIP5P shows 70\%.
  Then, the AIMD at 345~K is compared with TIP5P at 275~K (-20\%), 
where both distributions are, quite remarkably, hardly distinguishable.

\begin{figure}[t!]
\includegraphics*[scale=1.5]{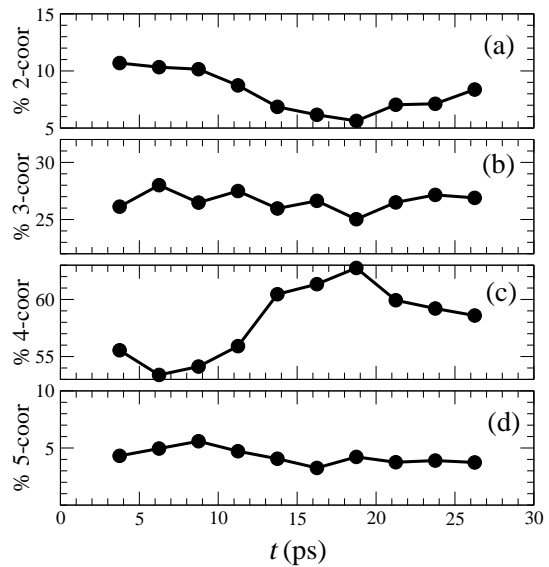}
\caption[]{Evolution of the proportion of molecules coordinated by two (a),
three (b), four (c), and five (d) molecules. Averages done in 7.5-ps-windows 
every 2.5 ps (simulation 2).}
\label{network-time}
\end{figure}

\begin{figure}[t!]
\includegraphics*[scale=1.3]{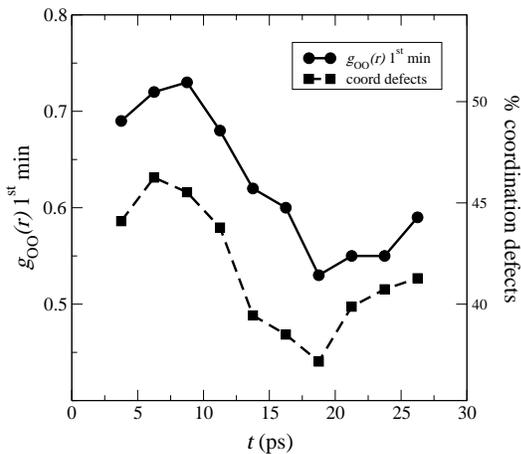}
\caption[]{Evolution of the first minimum of $g_{\rm OO}(r)$ and
the concentration of network defects, measured as \% of molecules
not tetra-coordinated (simulation 2).}
\label{correl}
\end{figure}

  Fig.~\ref{network-time} shows the evolution of the concentrations of
molecules with different coordinations.
  For consistency we have used the same ``moving window" approach as
before.
  The figure shows that the relaxation process is clearly associated 
to changes in defect concentrations, and thus to a reorganization of 
the HB network.  
  Furthermore, the non-equilibrium evolution of such defect concentrations
(Fig.~\ref{network-time}) is remarkably correlated with the evolution 
of the structural properties shown in Fig.~\ref{evolution}.
  This correlated behavior is further displayed in Fig.~\ref{correl}
where the dynamics of both properties are directly compared.
  Increases of four-coordinated molecules, with the consequent decrease
in coordination defects, result in a reduction of the diffusivity and
an enhancement of the structure in the RDFs.
  The under-coordinated molecules are the ones varying most, especially
the bi-coordinated.

  We have performed the same analysis for every simulation presented in
this work.
  The general trend is as described, namely that increases in the 
concentration of under-coordinated (particularly bi-coordinated) network 
defects correlate with an increasing diffusivity of the system, 
confirming the link between diffusivity and network imperfection.
  The same link was found in Ref.~\onlinecite{Giovambattista02}, where
the slow structural component of motion in supercooled water
was associated to transitions between basins in the potential
energy landscape. 
  These transitions occur through changes in the local structure of the 
HB network.

\begin{figure}[t!]
\includegraphics*[scale=1.5]{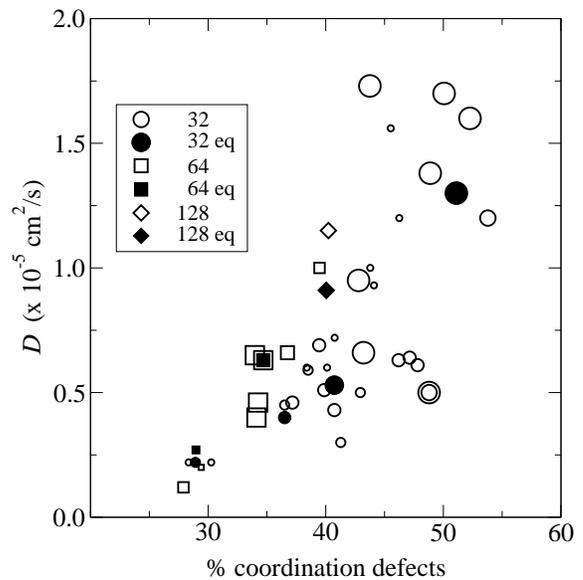}
\caption[]{Coordination defects versus diffusion coefficient ($D$) 
for all simulations. Open symbols represent non-equilibrium states 
as obtained in 7.5-ps-windows every 2.5 ps. Filled symbols represent 
equilibrium. Circles, squares and diamonds are for 32, 64, and 128
molecules, respectively. The size of each symbol reflects the 
(instantaneous) temperature, in five sizes, the smallest being for 
the range 295~K -- 305~K, and the largest for 335~K -- 345~K.}
\label{NDT}
\end{figure}

  Remarkably, the curve traced by a simulation in the diffusivity vs 
under-coordination plane is quite close and parallel to the 
equilibrium curve, obtained from joining the equilibrium points 
for the different simulations at different temperatures,
as shown in Fig.~\ref{NDT}.
  It is as if the diffusivity were mainly determined by the 
state of the network, the actual temperature becoming secondary.
  It is important to note as well that the state of the network is
not completely characterized by just the concentration of coordination 
defects.
  The spatial distribution of such defects will also be relevant.
  This will be explored elswhere.

  The evolution of the temperature along the simulations is 
consistent with the picture.
  When, for an initial temperature, the network is under-structured, 
the evolution toward more structuring (as in Fig.~\ref{evolution})
is accompanied by increasing temperature, i.e., the network is finding
regions of configuration space with lower potential energy.
  This is the situation when starting from empirical simulations
equilibrated at the target AIMD temperature, since force fields tend
to produce less structured liquids.
  If starting from an over-structured network, however, the slow
increase in network defects is accompanied by a decrease in system
temperature as the new defects are created.
  This is the case for simulation 3, for which the starting point was 
a previous AIMD simulation equilibrated at a lower temperature.

% --------------------

\section{Conclusions}

  We have carried out a series of AIMD simulations of liquid water
using the {\sc Siesta} method, for temperatures between 300K and 350K.
 The conclusions of the present study can be summarized as follows:

  $(i)$ The differences between the {\sc Siesta} method at the level
used in this study and PW-based AIMD methods\cite{Grossman04,Asthagiri03}
are less significant than the deviations between AIMD and experiment.
  This points to the more fundamental approximations (neglect of
quantum fluctuations in the dynamics, and of dispersion forces) as
responsible for the latter deviations.

  $(ii)$ The additional localization approximation imposed by the
the linear-scaling solver produces errors much less significant than
the ones mentioned above.
  This opens very good prospects for the study of complex systems in
interaction with water.

  $(iii)$ The comparison of different system sizes (32, 64 and 128
molecules) shows marginal size effects for both RDF and diffusivity.
 The largest system seems to show a slightly higher diffusivity,
to be confirmed by longer simulation times.

  $(iv)$ The AIMD results for RDF and diffusivity at a given temperature
compare well with the experimental results for a temperature 20\% lower, 
in fundamental agreement with PW AIMD results,\cite{Schwegler04} which
require a lowering of around 28\%.
  It means that the AIMD simulations performed in the past at
room temperature were in fact describing supercooled water, with
an effective temperature of around 240~K (at least for diffusivity and
RDF purposes).
  The 20\% temperature rescaling works also remarkably well for
the comparison between AIMD and TIP5P in the HB network imperfection.
  Besides its fundamental implications, a direct practical consequence of 
this conclusion is that, if an AIMD simulation at temperature $T$ is to be 
prepared by running an empirical model beforehand, it will be much more 
efficient to equilibrate the model to 0.8$T$ rather than $T$.

  $(v)$ A slow equilibration process of a time scale of at least
20 ps at AIMD's room temperature has been identified. 
  If it is related to the structural relaxation times characterized
experimentally,\cite{Sette04} the process could have a much shorter
time scale ($\approx$1~ps) at {\it real} room temperature, {\it i.e.},
for our AIMD simulations scaled up in temperature, or for AIMD
with a better performing GGA.

  $(vi)$ In this equilibration process, the ``instantaneous" diffusivity 
correlates with some instantaneous structural properties captured in the 
RDF, much more than with the actual instantaneous temperature.

  $(vii)$ HB network rearrangements have been proposed to be
behind the slow equilibration process.
  Network imperfection (mainly the proportion of under-coordinated
molecules) has been found to correlate very strongly with
the diffusivity and the RDFs in their non-equilibrium evolution.

  These last findings have important implications in the way we see
the present DFT problems in the description of liquid water.
  Rather than being related to overestimation of energy barriers,
this work points to an overestimation of the energy of formation
of coordination defects.
  It is, therefore, not so much about transition states
in the energy landscape as about the energy difference between
basins in that landscape. 
  Finally, we are convinced that the direct link described here between 
diffusivity and network imperfection is a property of liquid water,
not an artefact of DFT.\cite{Giovambattista02} 

\acknowledgments

  We thank M. Sprik and J. VandeVondele for useful discussions,
and M. C. Payne for his suggestion to consider temperature scaling.
  We acknowledge financial support from the British Engineering and
Physical Sciences Research Council, the Cambridge European Trust,
and the Comunidad Aut\'onoma de Madrid.
  The calculations were performed in the Cambridge Cranfield 
High Performance Computing Facility.

\bibliographystyle{apsrev}
\bibliography{mvf} 

\begin{thebibliography}{57}
\expandafter\ifx\csname natexlab\endcsname\relax\def\natexlab#1{#1}\fi
\expandafter\ifx\csname bibnamefont\endcsname\relax
  \def\bibnamefont#1{#1}\fi
\expandafter\ifx\csname bibfnamefont\endcsname\relax
  \def\bibfnamefont#1{#1}\fi
\expandafter\ifx\csname citenamefont\endcsname\relax
  \def\citenamefont#1{#1}\fi
\expandafter\ifx\csname url\endcsname\relax
  \def\url#1{\texttt{#1}}\fi
\expandafter\ifx\csname urlprefix\endcsname\relax\def\urlprefix{URL }\fi
\providecommand{\bibinfo}[2]{#2}
\providecommand{\eprint}[2][]{\url{#2}}

\bibitem[{\citenamefont{Sprik et~al.}(1996)\citenamefont{Sprik, Hutter, and
  Parrinello}}]{Sprik96}
\bibinfo{author}{\bibfnamefont{M.}~\bibnamefont{Sprik}},
  \bibinfo{author}{\bibfnamefont{J.}~\bibnamefont{Hutter}}, \bibnamefont{and}
  \bibinfo{author}{\bibfnamefont{M.}~\bibnamefont{Parrinello}},
  \bibinfo{journal}{J. Chem. Phys.} \textbf{\bibinfo{volume}{105}},
  \bibinfo{pages}{1142} (\bibinfo{year}{1996}).

\bibitem[{\citenamefont{Ortega et~al.}(1996)\citenamefont{Ortega, Lewis, and
  Sankey}}]{Ortega96}
\bibinfo{author}{\bibfnamefont{J.}~\bibnamefont{Ortega}},
  \bibinfo{author}{\bibfnamefont{J.~P.} \bibnamefont{Lewis}}, \bibnamefont{and}
  \bibinfo{author}{\bibfnamefont{O.}~\bibnamefont{Sankey}},
  \bibinfo{journal}{J. Chem. Phys.} \textbf{\bibinfo{volume}{106}},
  \bibinfo{pages}{3696} (\bibinfo{year}{1996}).

\bibitem[{\citenamefont{Silvestrelli and
  Parrinello}(1999{\natexlab{a}})}]{Silvestrelli99jcp}
\bibinfo{author}{\bibfnamefont{P.~L.} \bibnamefont{Silvestrelli}}
  \bibnamefont{and}
  \bibinfo{author}{\bibfnamefont{M.}~\bibnamefont{Parrinello}},
  \bibinfo{journal}{J. Chem. Phys.} \textbf{\bibinfo{volume}{111}},
  \bibinfo{pages}{3572} (\bibinfo{year}{1999}{\natexlab{a}}).

\bibitem[{\citenamefont{Silvestrelli and
  Parrinello}(1999{\natexlab{b}})}]{Silvestrelli99prl}
\bibinfo{author}{\bibfnamefont{P.~L.} \bibnamefont{Silvestrelli}}
  \bibnamefont{and}
  \bibinfo{author}{\bibfnamefont{M.}~\bibnamefont{Parrinello}},
  \bibinfo{journal}{Phys. Rev. Lett.} \textbf{\bibinfo{volume}{82}},
  \bibinfo{pages}{3308} (\bibinfo{year}{1999}{\natexlab{b}}).

\bibitem[{\citenamefont{Geissler et~al.}(2001)\citenamefont{Geissler, Dellago,
  Chandler, Hutter, and Parrinello}}]{parrinelloOH}
\bibinfo{author}{\bibfnamefont{P.~L.} \bibnamefont{Geissler}},
  \bibinfo{author}{\bibfnamefont{C.}~\bibnamefont{Dellago}},
  \bibinfo{author}{\bibfnamefont{D.}~\bibnamefont{Chandler}},
  \bibinfo{author}{\bibfnamefont{J.}~\bibnamefont{Hutter}}, \bibnamefont{and}
  \bibinfo{author}{\bibfnamefont{M.}~\bibnamefont{Parrinello}},
  \bibinfo{journal}{Science} \textbf{\bibinfo{volume}{291}},
  \bibinfo{pages}{2121} (\bibinfo{year}{2001}).

\bibitem[{\citenamefont{Boero et~al.}(2003)\citenamefont{Boero, Parrinello,
  Terakura, Ikeshoji, and Liew}}]{parrinello-wet-e}
\bibinfo{author}{\bibfnamefont{M.}~\bibnamefont{Boero}},
  \bibinfo{author}{\bibfnamefont{M.}~\bibnamefont{Parrinello}},
  \bibinfo{author}{\bibfnamefont{K.}~\bibnamefont{Terakura}},
  \bibinfo{author}{\bibfnamefont{T.}~\bibnamefont{Ikeshoji}}, \bibnamefont{and}
  \bibinfo{author}{\bibfnamefont{C.~C.} \bibnamefont{Liew}},
  \bibinfo{journal}{Phys. Rev. Lett.} \textbf{\bibinfo{volume}{90}},
  \bibinfo{pages}{226403} (\bibinfo{year}{2003}).

\bibitem[{\citenamefont{Izvekov and Voth}(2002)}]{Voth02}
\bibinfo{author}{\bibfnamefont{S.}~\bibnamefont{Izvekov}} \bibnamefont{and}
  \bibinfo{author}{\bibfnamefont{G.~A.} \bibnamefont{Voth}},
  \bibinfo{journal}{J. Chem. Phys.} \textbf{\bibinfo{volume}{116}},
  \bibinfo{pages}{10372} (\bibinfo{year}{2002}).

\bibitem[{\citenamefont{Grossman et~al.}(2004)\citenamefont{Grossman,
  Schwegler, Draeger, Gygi, and Galli}}]{Grossman04}
\bibinfo{author}{\bibfnamefont{J.~C.} \bibnamefont{Grossman}},
  \bibinfo{author}{\bibfnamefont{E.}~\bibnamefont{Schwegler}},
  \bibinfo{author}{\bibfnamefont{E.~W.} \bibnamefont{Draeger}},
  \bibinfo{author}{\bibfnamefont{F.}~\bibnamefont{Gygi}}, \bibnamefont{and}
  \bibinfo{author}{\bibfnamefont{G.}~\bibnamefont{Galli}}, \bibinfo{journal}{J.
  Chem. Phys.} \textbf{\bibinfo{volume}{120}}, \bibinfo{pages}{300}
  (\bibinfo{year}{2004}).

\bibitem[{\citenamefont{Schwegler et~al.}(2004)\citenamefont{Schwegler,
  Grossman, Gygi, and Galli}}]{Schwegler04}
\bibinfo{author}{\bibfnamefont{E.}~\bibnamefont{Schwegler}},
  \bibinfo{author}{\bibfnamefont{J.~C.} \bibnamefont{Grossman}},
  \bibinfo{author}{\bibfnamefont{F.}~\bibnamefont{Gygi}}, \bibnamefont{and}
  \bibinfo{author}{\bibfnamefont{G.}~\bibnamefont{Galli}},
  \bibinfo{journal}{arXiv:cond-mat} p. \bibinfo{pages}{0405561}
  (\bibinfo{year}{2004}).

\bibitem[{\citenamefont{Asthagiri et~al.}(2003)\citenamefont{Asthagiri, Pratt,
  and Kress}}]{Asthagiri03}
\bibinfo{author}{\bibfnamefont{D.}~\bibnamefont{Asthagiri}},
  \bibinfo{author}{\bibfnamefont{L.~R.} \bibnamefont{Pratt}}, \bibnamefont{and}
  \bibinfo{author}{\bibfnamefont{J.~D.} \bibnamefont{Kress}},
  \bibinfo{journal}{Phys. Rev. E} \textbf{\bibinfo{volume}{68}},
  \bibinfo{pages}{41505} (\bibinfo{year}{2003}).

\bibitem[{\citenamefont{Mills}(1973)}]{waterexpD}
\bibinfo{author}{\bibfnamefont{R.}~\bibnamefont{Mills}}, \bibinfo{journal}{J.
  Chem. Phys.} \textbf{\bibinfo{volume}{77}}, \bibinfo{pages}{685}
  (\bibinfo{year}{1973}).

\bibitem[{\citenamefont{Soper}(2000)}]{waterexp1}
\bibinfo{author}{\bibfnamefont{A.~K.} \bibnamefont{Soper}},
  \bibinfo{journal}{Chem. Phys.} \textbf{\bibinfo{volume}{258}},
  \bibinfo{pages}{121} (\bibinfo{year}{2000}).

\bibitem[{\citenamefont{Sorenson et~al.}(2000)\citenamefont{Sorenson, Hura,
  Glaeser, and Head-Gordon}}]{waterexp2}
\bibinfo{author}{\bibfnamefont{J.~M.} \bibnamefont{Sorenson}},
  \bibinfo{author}{\bibfnamefont{G.}~\bibnamefont{Hura}},
  \bibinfo{author}{\bibfnamefont{R.~M.} \bibnamefont{Glaeser}},
  \bibnamefont{and}
  \bibinfo{author}{\bibfnamefont{T.}~\bibnamefont{Head-Gordon}},
  \bibinfo{journal}{J. Chem. Phys.} \textbf{\bibinfo{volume}{113}},
  \bibinfo{pages}{9149} (\bibinfo{year}{2000}).

\bibitem[{\citenamefont{Kohn et~al.}(1998)\citenamefont{Kohn, Meir, and
  Makarov}}]{Kohn98}
\bibinfo{author}{\bibfnamefont{W.}~\bibnamefont{Kohn}},
  \bibinfo{author}{\bibfnamefont{Y.}~\bibnamefont{Meir}}, \bibnamefont{and}
  \bibinfo{author}{\bibfnamefont{D.~E.} \bibnamefont{Makarov}},
  \bibinfo{journal}{Phys. Rev. Lett.} \textbf{\bibinfo{volume}{80}},
  \bibinfo{pages}{4153} (\bibinfo{year}{1998}).

\bibitem[{\citenamefont{Dion et~al.}(2004)\citenamefont{Dion, Rydberg,
  Schröder, Langreth, and Lundqvist}}]{Lundqvist04}
\bibinfo{author}{\bibfnamefont{M.}~\bibnamefont{Dion}},
  \bibinfo{author}{\bibfnamefont{H.}~\bibnamefont{Rydberg}},
  \bibinfo{author}{\bibfnamefont{E.}~\bibnamefont{Schröder}},
  \bibinfo{author}{\bibfnamefont{D.~C.} \bibnamefont{Langreth}},
  \bibnamefont{and} \bibinfo{author}{\bibfnamefont{B.~I.}
  \bibnamefont{Lundqvist}}, \bibinfo{journal}{Phys. Rev. Lett.}
  \textbf{\bibinfo{volume}{92}}, \bibinfo{pages}{246401}
  (\bibinfo{year}{2004}).

\bibitem[{\citenamefont{Chen et~al.}(2003)\citenamefont{Chen, Ivanov, Klein,
  and Parrinello}}]{parrinelloqtm}
\bibinfo{author}{\bibfnamefont{B.}~\bibnamefont{Chen}},
  \bibinfo{author}{\bibfnamefont{I.}~\bibnamefont{Ivanov}},
  \bibinfo{author}{\bibfnamefont{M.~L.} \bibnamefont{Klein}}, \bibnamefont{and}
  \bibinfo{author}{\bibfnamefont{M.}~\bibnamefont{Parrinello}},
  \bibinfo{journal}{Phys. Rev. Lett.} \textbf{\bibinfo{volume}{91}},
  \bibinfo{pages}{215503} (\bibinfo{year}{2003}).

\bibitem[{\citenamefont{Raugei and Klein}(2003)}]{kleinqtm}
\bibinfo{author}{\bibfnamefont{S.}~\bibnamefont{Raugei}} \bibnamefont{and}
  \bibinfo{author}{\bibfnamefont{M.~L.} \bibnamefont{Klein}},
  \bibinfo{journal}{J. Am. Chem. Soc.} \textbf{\bibinfo{volume}{125}},
  \bibinfo{pages}{8992} (\bibinfo{year}{2003}).

\bibitem[{\citenamefont{Ordej\'on et~al.}(1996)\citenamefont{Ordej\'on,
  Artacho, and Soler}}]{SiestaPRBRC}
\bibinfo{author}{\bibfnamefont{P.}~\bibnamefont{Ordej\'on}},
  \bibinfo{author}{\bibfnamefont{E.}~\bibnamefont{Artacho}}, \bibnamefont{and}
  \bibinfo{author}{\bibfnamefont{J.~M.} \bibnamefont{Soler}},
  \bibinfo{journal}{Phys. Rev. B} \textbf{\bibinfo{volume}{53}},
  \bibinfo{pages}{10441} (\bibinfo{year}{1996}).

\bibitem[{\citenamefont{Soler et~al.}(2002)\citenamefont{Soler, Artacho, Gale,
  Garc\'{\i}a, Junquera, Ordej\'on, and S\'anchez-Portal}}]{SiestaJPCM}
\bibinfo{author}{\bibfnamefont{J.~M.} \bibnamefont{Soler}},
  \bibinfo{author}{\bibfnamefont{E.}~\bibnamefont{Artacho}},
  \bibinfo{author}{\bibfnamefont{J.~D.} \bibnamefont{Gale}},
  \bibinfo{author}{\bibfnamefont{A.}~\bibnamefont{Garc\'{\i}a}},
  \bibinfo{author}{\bibfnamefont{J.}~\bibnamefont{Junquera}},
  \bibinfo{author}{\bibfnamefont{P.}~\bibnamefont{Ordej\'on}},
  \bibnamefont{and}
  \bibinfo{author}{\bibfnamefont{D.}~\bibnamefont{S\'anchez-Portal}},
  \bibinfo{journal}{J Phys. Condens. Matter} \textbf{\bibinfo{volume}{14}},
  \bibinfo{pages}{2745} (\bibinfo{year}{2002}).

\bibitem[{\citenamefont{Ordej\'on et~al.}(1995)\citenamefont{Ordej\'on,
  Drabold, Grumbach, and Martin}}]{Ordejon95}
\bibinfo{author}{\bibfnamefont{P.}~\bibnamefont{Ordej\'on}},
  \bibinfo{author}{\bibfnamefont{A.}~\bibnamefont{Drabold}},
  \bibinfo{author}{\bibfnamefont{M.~P.} \bibnamefont{Grumbach}},
  \bibnamefont{and} \bibinfo{author}{\bibfnamefont{R.~M.}
  \bibnamefont{Martin}}, \bibinfo{journal}{Phys. Rev. B}
  \textbf{\bibinfo{volume}{51}}, \bibinfo{pages}{1456} (\bibinfo{year}{1995}).

\bibitem[{\citenamefont{Kim et~al.}(1995)\citenamefont{Kim, Mauri, and
  Galli}}]{Kim-mauri-galli}
\bibinfo{author}{\bibfnamefont{J.}~\bibnamefont{Kim}},
  \bibinfo{author}{\bibfnamefont{F.}~\bibnamefont{Mauri}}, \bibnamefont{and}
  \bibinfo{author}{\bibfnamefont{G.}~\bibnamefont{Galli}},
  \bibinfo{journal}{Phys. Rev. B} \textbf{\bibinfo{volume}{52}},
  \bibinfo{pages}{1640} (\bibinfo{year}{1995}).

\bibitem[{\citenamefont{Kohn and Sham}(1965)}]{Kohn-Sham}
\bibinfo{author}{\bibfnamefont{W.}~\bibnamefont{Kohn}} \bibnamefont{and}
  \bibinfo{author}{\bibfnamefont{L.~J.} \bibnamefont{Sham}},
  \bibinfo{journal}{Phys. Rev.} \textbf{\bibinfo{volume}{140}},
  \bibinfo{pages}{a1133} (\bibinfo{year}{1965}).

\bibitem[{\citenamefont{Hohenberg and Kohn}(1964)}]{Hohenberg-K}
\bibinfo{author}{\bibfnamefont{P.}~\bibnamefont{Hohenberg}} \bibnamefont{and}
  \bibinfo{author}{\bibfnamefont{W.}~\bibnamefont{Kohn}},
  \bibinfo{journal}{Phys. Rev.} \textbf{\bibinfo{volume}{136}},
  \bibinfo{pages}{b864} (\bibinfo{year}{1964}).

\bibitem[{\citenamefont{Becke}(1998)}]{blyp1}
\bibinfo{author}{\bibfnamefont{A.~D.} \bibnamefont{Becke}},
  \bibinfo{journal}{Phys. Rev. A} \textbf{\bibinfo{volume}{38}},
  \bibinfo{pages}{3098} (\bibinfo{year}{1998}).

\bibitem[{\citenamefont{Lee et~al.}(1988)\citenamefont{Lee, Yang, and
  Parr}}]{blyp2}
\bibinfo{author}{\bibfnamefont{C.}~\bibnamefont{Lee}},
  \bibinfo{author}{\bibfnamefont{W.}~\bibnamefont{Yang}}, \bibnamefont{and}
  \bibinfo{author}{\bibfnamefont{R.~G.} \bibnamefont{Parr}},
  \bibinfo{journal}{Phys. Rev. B} \textbf{\bibinfo{volume}{37}},
  \bibinfo{pages}{785} (\bibinfo{year}{1988}).

\bibitem[{\citenamefont{Troullier and Martins}(1991)}]{Troullier-Martins}
\bibinfo{author}{\bibfnamefont{N.}~\bibnamefont{Troullier}} \bibnamefont{and}
  \bibinfo{author}{\bibfnamefont{J.~L.} \bibnamefont{Martins}},
  \bibinfo{journal}{Phys. Rev. B} \textbf{\bibinfo{volume}{43}},
  \bibinfo{pages}{1993} (\bibinfo{year}{1991}).

\bibitem[{inp()}]{inp}
\bibinfo{note}{The pseudopotential cutoff radii were 1.15$a_0$ for all channels
  in O, and 1.20$a_0$ in H. For O, a non-linear partial-core
  correction\cite{pcec} was used in order to ensure a smooth GGA
  pseudopotential near the nucleus.\cite{Hamann} It had the form proposed by M.
  M. G. Alemany and J. L. Martins (unpublished) $pc(r)=e^{\alpha+\beta r^2 +
  \gamma r^4}$, and a matching radius of 1.17$a_0$.}

\bibitem[{\citenamefont{Kleinman and Bylander}(1982)}]{Kleinman-Bylander}
\bibinfo{author}{\bibfnamefont{L.}~\bibnamefont{Kleinman}} \bibnamefont{and}
  \bibinfo{author}{\bibfnamefont{D.~M.} \bibnamefont{Bylander}},
  \bibinfo{journal}{Phys. Rev. Lett.} \textbf{\bibinfo{volume}{48}},
  \bibinfo{pages}{1425} (\bibinfo{year}{1982}).

\bibitem[{\citenamefont{Junquera et~al.}(2001)\citenamefont{Junquera, Paz,
  S\'anchez-Portal, and Artacho}}]{javibases}
\bibinfo{author}{\bibfnamefont{J.}~\bibnamefont{Junquera}},
  \bibinfo{author}{\bibfnamefont{O.}~\bibnamefont{Paz}},
  \bibinfo{author}{\bibfnamefont{D.}~\bibnamefont{S\'anchez-Portal}},
  \bibnamefont{and} \bibinfo{author}{\bibfnamefont{E.}~\bibnamefont{Artacho}},
  \bibinfo{journal}{Phys. Rev. B} \textbf{\bibinfo{volume}{64}},
  \bibinfo{pages}{235111} (\bibinfo{year}{2001}).

\bibitem[{\citenamefont{Anglada et~al.}(2002)\citenamefont{Anglada, Soler,
  Junquera, and Artacho}}]{edubases}
\bibinfo{author}{\bibfnamefont{E.}~\bibnamefont{Anglada}},
  \bibinfo{author}{\bibfnamefont{J.~M.} \bibnamefont{Soler}},
  \bibinfo{author}{\bibfnamefont{J.}~\bibnamefont{Junquera}}, \bibnamefont{and}
  \bibinfo{author}{\bibfnamefont{E.}~\bibnamefont{Artacho}},
  \bibinfo{journal}{Phys. Rev. B} \textbf{\bibinfo{volume}{66}},
  \bibinfo{pages}{205101} (\bibinfo{year}{2002}).

\bibitem[{\citenamefont{Benedict et~al.}(1975)\citenamefont{Benedict, Gailar,
  and Plyler}}]{wbenedict75}
\bibinfo{author}{\bibfnamefont{W.~S.} \bibnamefont{Benedict}},
  \bibinfo{author}{\bibfnamefont{N.}~\bibnamefont{Gailar}}, \bibnamefont{and}
  \bibinfo{author}{\bibfnamefont{E.~K.} \bibnamefont{Plyler}},
  \bibinfo{journal}{J. Chem. Phys.} \textbf{\bibinfo{volume}{24}},
  \bibinfo{pages}{1139} (\bibinfo{year}{1975}).

\bibitem[{\citenamefont{Clough et~al.}(1973)\citenamefont{Clough, Beers, Klein,
  and Rothman}}]{sclough73}
\bibinfo{author}{\bibfnamefont{S.~A.} \bibnamefont{Clough}},
  \bibinfo{author}{\bibfnamefont{Y.}~\bibnamefont{Beers}},
  \bibinfo{author}{\bibfnamefont{G.~P.} \bibnamefont{Klein}}, \bibnamefont{and}
  \bibinfo{author}{\bibfnamefont{L.~S.} \bibnamefont{Rothman}},
  \bibinfo{journal}{J. Chem. Phys.} \textbf{\bibinfo{volume}{59}},
  \bibinfo{pages}{2254} (\bibinfo{year}{1973}).

\bibitem[{\citenamefont{Odutola and Dyke}(1980)}]{jodutola80}
\bibinfo{author}{\bibfnamefont{J.~A.} \bibnamefont{Odutola}} \bibnamefont{and}
  \bibinfo{author}{\bibfnamefont{T.~R.} \bibnamefont{Dyke}},
  \bibinfo{journal}{J. Chem. Phys.} \textbf{\bibinfo{volume}{72}},
  \bibinfo{pages}{5062} (\bibinfo{year}{1980}).

\bibitem[{\citenamefont{Curtiss et~al.}(1979)\citenamefont{Curtiss, Frurip, and
  Blander}}]{lcurtiss79}
\bibinfo{author}{\bibfnamefont{L.~A.} \bibnamefont{Curtiss}},
  \bibinfo{author}{\bibfnamefont{D.~J.} \bibnamefont{Frurip}},
  \bibnamefont{and} \bibinfo{author}{\bibfnamefont{M.}~\bibnamefont{Blander}},
  \bibinfo{journal}{J. Chem. Phys.} \textbf{\bibinfo{volume}{71}},
  \bibinfo{pages}{2703} (\bibinfo{year}{1979}).

\bibitem[{\citenamefont{Allen and Tildesley}(1987)}]{allen-tildesley}
\bibinfo{author}{\bibfnamefont{M.~P.} \bibnamefont{Allen}} \bibnamefont{and}
  \bibinfo{author}{\bibfnamefont{D.~J.} \bibnamefont{Tildesley}},
  \emph{\bibinfo{title}{{``Computer simulation of Liquids"}}}
  (\bibinfo{publisher}{Oxford University Press}, \bibinfo{year}{1987}).

\bibitem[{\citenamefont{Vaisman et~al.}(1993)\citenamefont{Vaisman, Perea, and
  Berkowitz}}]{Vaisman93}
\bibinfo{author}{\bibfnamefont{I.}~\bibnamefont{Vaisman}},
  \bibinfo{author}{\bibfnamefont{L.}~\bibnamefont{Perea}}, \bibnamefont{and}
  \bibinfo{author}{\bibfnamefont{M.~L.} \bibnamefont{Berkowitz}},
  \bibinfo{journal}{J. Chem. Phys.} \textbf{\bibinfo{volume}{98}},
  \bibinfo{pages}{9859} (\bibinfo{year}{1993}).

\bibitem[{\citenamefont{Mahoney and Jorgensen}(2000)}]{tip5p}
\bibinfo{author}{\bibfnamefont{M.~W.} \bibnamefont{Mahoney}} \bibnamefont{and}
  \bibinfo{author}{\bibfnamefont{W.~L.} \bibnamefont{Jorgensen}},
  \bibinfo{journal}{J. Chem. Phys.} \textbf{\bibinfo{volume}{112}},
  \bibinfo{pages}{8910} (\bibinfo{year}{2000}).

\bibitem[{\citenamefont{Berendsen et~al.}(1987)\citenamefont{Berendsen,
  Grigera, and Straatsma}}]{SPC/E}
\bibinfo{author}{\bibfnamefont{H.~J.~C.} \bibnamefont{Berendsen}},
  \bibinfo{author}{\bibfnamefont{J.~R.} \bibnamefont{Grigera}},
  \bibnamefont{and} \bibinfo{author}{\bibfnamefont{T.~P.}
  \bibnamefont{Straatsma}}, \bibinfo{journal}{J. Chem. Phys.}
  \textbf{\bibinfo{volume}{91}}, \bibinfo{pages}{6269} (\bibinfo{year}{1987}).

\bibitem[{\citenamefont{Berendsen et~al.}(1995)\citenamefont{Berendsen, H.J.C.,
  van~der Spoel, D., and van Drunen}}]{gromacs1}
\bibinfo{author}{\bibnamefont{Berendsen}},
  \bibinfo{author}{\bibnamefont{H.J.C.}}, \bibinfo{author}{\bibnamefont{van~der
  Spoel}}, \bibinfo{author}{\bibnamefont{D.}}, \bibnamefont{and}
  \bibinfo{author}{\bibfnamefont{R.}~\bibnamefont{van Drunen}},
  \bibinfo{journal}{Comp. Phys. Comm.} \textbf{\bibinfo{volume}{91}},
  \bibinfo{pages}{43} (\bibinfo{year}{1995}).

\bibitem[{\citenamefont{Lindahl et~al.}(2001)\citenamefont{Lindahl, E., Hess,
  B., and van~der Spoel}}]{gromacs2}
\bibinfo{author}{\bibnamefont{Lindahl}}, \bibinfo{author}{\bibnamefont{E.}},
  \bibinfo{author}{\bibnamefont{Hess}}, \bibinfo{author}{\bibnamefont{B.}},
  \bibnamefont{and} \bibinfo{author}{\bibfnamefont{D.}~\bibnamefont{van~der
  Spoel}}, \bibinfo{journal}{J. Mol. Mod.} \textbf{\bibinfo{volume}{7}},
  \bibinfo{pages}{306} (\bibinfo{year}{2001}).

\bibitem[{\citenamefont{Berendsen et~al.}(1984)\citenamefont{Berendsen, Pstman,
  van Gunsteren, DiNola, and Haak}}]{berendsen}
\bibinfo{author}{\bibfnamefont{H.~J.~C.} \bibnamefont{Berendsen}},
  \bibinfo{author}{\bibfnamefont{J.~P.~M.} \bibnamefont{Pstman}},
  \bibinfo{author}{\bibfnamefont{W.~F.} \bibnamefont{van Gunsteren}},
  \bibinfo{author}{\bibfnamefont{A.}~\bibnamefont{DiNola}}, \bibnamefont{and}
  \bibinfo{author}{\bibfnamefont{J.~R.} \bibnamefont{Haak}},
  \bibinfo{journal}{J. Chem. Phys.} \textbf{\bibinfo{volume}{81}},
  \bibinfo{pages}{3684} (\bibinfo{year}{1984}).

\bibitem[{\citenamefont{Mahoney and Jorgensen}(2001)}]{tip5pDif}
\bibinfo{author}{\bibfnamefont{M.~W.} \bibnamefont{Mahoney}} \bibnamefont{and}
  \bibinfo{author}{\bibfnamefont{W.~L.} \bibnamefont{Jorgensen}},
  \bibinfo{journal}{J. Chem. Phys.} \textbf{\bibinfo{volume}{114}},
  \bibinfo{pages}{363} (\bibinfo{year}{2001}).

\bibitem[{\citenamefont{Perdew et~al.}(1996)\citenamefont{Perdew, Burke, and
  Ernzerhof}}]{PBE}
\bibinfo{author}{\bibfnamefont{J.~P.} \bibnamefont{Perdew}},
  \bibinfo{author}{\bibfnamefont{K.}~\bibnamefont{Burke}}, \bibnamefont{and}
  \bibinfo{author}{\bibfnamefont{M.}~\bibnamefont{Ernzerhof}},
  \bibinfo{journal}{Phys. Rev. Lett.} \textbf{\bibinfo{volume}{77}},
  \bibinfo{pages}{3865} (\bibinfo{year}{1996}).

\bibitem[{\citenamefont{Hammer et~al.}(1999)\citenamefont{Hammer, Hansen, and
  Norskov}}]{rpbe}
\bibinfo{author}{\bibfnamefont{B.}~\bibnamefont{Hammer}},
  \bibinfo{author}{\bibfnamefont{L.~B.} \bibnamefont{Hansen}},
  \bibnamefont{and} \bibinfo{author}{\bibfnamefont{J.~K.}
  \bibnamefont{Norskov}}, \bibinfo{journal}{Phys. Rev. B}
  \textbf{\bibinfo{volume}{59}}, \bibinfo{pages}{7413} (\bibinfo{year}{1999}).

\bibitem[{\citenamefont{Masciovecchio et~al.}(2004)\citenamefont{Masciovecchio,
  Santucci, Gessini, Fonzo, Ruocco, and Sette}}]{Sette04}
\bibinfo{author}{\bibfnamefont{C.}~\bibnamefont{Masciovecchio}},
  \bibinfo{author}{\bibfnamefont{S.~C.} \bibnamefont{Santucci}},
  \bibinfo{author}{\bibfnamefont{A.}~\bibnamefont{Gessini}},
  \bibinfo{author}{\bibfnamefont{S.~D.} \bibnamefont{Fonzo}},
  \bibinfo{author}{\bibfnamefont{G.}~\bibnamefont{Ruocco}}, \bibnamefont{and}
  \bibinfo{author}{\bibfnamefont{F.}~\bibnamefont{Sette}},
  \bibinfo{journal}{Phys. Rev. Lett.} \textbf{\bibinfo{volume}{92}},
  \bibinfo{pages}{255507} (\bibinfo{year}{2004}).

\bibitem[{\citenamefont{Stillinger}(1980)}]{stillingerwater}
\bibinfo{author}{\bibfnamefont{F.}~\bibnamefont{Stillinger}},
  \bibinfo{journal}{Science} \textbf{\bibinfo{volume}{209}},
  \bibinfo{pages}{451} (\bibinfo{year}{1980}).

\bibitem[{\citenamefont{Guo et~al.}(2002)\citenamefont{Guo, Luo, Augustsson,
  Rubensson, Sathe, Agren, siegbahn, and Nordgren}}]{Guo02}
\bibinfo{author}{\bibfnamefont{J.-H.} \bibnamefont{Guo}},
  \bibinfo{author}{\bibfnamefont{Y.}~\bibnamefont{Luo}},
  \bibinfo{author}{\bibfnamefont{A.}~\bibnamefont{Augustsson}},
  \bibinfo{author}{\bibfnamefont{J.~E.} \bibnamefont{Rubensson}},
  \bibinfo{author}{\bibfnamefont{C.}~\bibnamefont{Sathe}},
  \bibinfo{author}{\bibfnamefont{H.}~\bibnamefont{Agren}},
  \bibinfo{author}{\bibfnamefont{H.}~\bibnamefont{siegbahn}}, \bibnamefont{and}
  \bibinfo{author}{\bibfnamefont{J.}~\bibnamefont{Nordgren}},
  \bibinfo{journal}{Phys. Rev. Lett.} \textbf{\bibinfo{volume}{89}},
  \bibinfo{pages}{137402} (\bibinfo{year}{2002}).

\bibitem[{\citenamefont{Romero et~al.}(2001)\citenamefont{Romero, Silvestrelli,
  and Parrinello}}]{Romero01}
\bibinfo{author}{\bibfnamefont{A.}~\bibnamefont{Romero}},
  \bibinfo{author}{\bibfnamefont{P.~L.} \bibnamefont{Silvestrelli}},
  \bibnamefont{and}
  \bibinfo{author}{\bibfnamefont{M.}~\bibnamefont{Parrinello}},
  \bibinfo{journal}{J. Chem. Phys.} \textbf{\bibinfo{volume}{115}},
  \bibinfo{pages}{115} (\bibinfo{year}{2001}).

\bibitem[{\citenamefont{Ghanty et~al.}(2000)\citenamefont{Ghanty, Staroverov,
  Koren, and Davidson}}]{Ghanty00}
\bibinfo{author}{\bibfnamefont{T.~K.} \bibnamefont{Ghanty}},
  \bibinfo{author}{\bibfnamefont{V.~N.} \bibnamefont{Staroverov}},
  \bibinfo{author}{\bibfnamefont{P.~K.} \bibnamefont{Koren}}, \bibnamefont{and}
  \bibinfo{author}{\bibfnamefont{E.~R.} \bibnamefont{Davidson}},
  \bibinfo{journal}{J. Am. Chem. Soc.} \textbf{\bibinfo{volume}{122}},
  \bibinfo{pages}{1210} (\bibinfo{year}{2000}).

\bibitem[{\citenamefont{Fern\'andez-Serra and Artacho}()}]{hbondpaper}
\bibinfo{author}{\bibfnamefont{M.~V.} \bibnamefont{Fern\'andez-Serra}}
  \bibnamefont{and} \bibinfo{author}{\bibfnamefont{E.}~\bibnamefont{Artacho}},
  \bibinfo{howpublished}{to be published}.

\bibitem[{\citenamefont{Luzar}(2000)}]{Luzar00}
\bibinfo{author}{\bibfnamefont{A.}~\bibnamefont{Luzar}}, \bibinfo{journal}{J.
  Chem. Phys.} \textbf{\bibinfo{volume}{113}}, \bibinfo{pages}{10663}
  (\bibinfo{year}{2000}).

\bibitem[{\citenamefont{Luzar and Chandler}(1996)}]{Chandler96}
\bibinfo{author}{\bibfnamefont{A.}~\bibnamefont{Luzar}} \bibnamefont{and}
  \bibinfo{author}{\bibfnamefont{D.}~\bibnamefont{Chandler}},
  \bibinfo{journal}{Nature} \textbf{\bibinfo{volume}{379}}, \bibinfo{pages}{55}
  (\bibinfo{year}{1996}).

\bibitem[{\citenamefont{Starr et~al.}(2000)\citenamefont{Starr, Nielsen, and
  Stanley}}]{Starr00}
\bibinfo{author}{\bibfnamefont{F.~W.} \bibnamefont{Starr}},
  \bibinfo{author}{\bibfnamefont{J.~K.} \bibnamefont{Nielsen}},
  \bibnamefont{and} \bibinfo{author}{\bibfnamefont{E.}~\bibnamefont{Stanley}},
  \bibinfo{journal}{Phys. Rev. E} \textbf{\bibinfo{volume}{62}},
  \bibinfo{pages}{579} (\bibinfo{year}{2000}).

\bibitem[{\citenamefont{Gallo et~al.}(1996)\citenamefont{Gallo, Sciortino,
  Tartaglia, and Chen}}]{Gallo96}
\bibinfo{author}{\bibfnamefont{P.}~\bibnamefont{Gallo}},
  \bibinfo{author}{\bibfnamefont{F.}~\bibnamefont{Sciortino}},
  \bibinfo{author}{\bibfnamefont{P.}~\bibnamefont{Tartaglia}},
  \bibnamefont{and} \bibinfo{author}{\bibfnamefont{S.-H.} \bibnamefont{Chen}},
  \bibinfo{journal}{Phys. Rev. Lett.} \textbf{\bibinfo{volume}{76}},
  \bibinfo{pages}{2730} (\bibinfo{year}{1996}).

\bibitem[{\citenamefont{Giovambattista
  et~al.}(2002)\citenamefont{Giovambattista, Starr, Sciortino, Buldyrev, and
  Stanley}}]{Giovambattista02}
\bibinfo{author}{\bibfnamefont{N.}~\bibnamefont{Giovambattista}},
  \bibinfo{author}{\bibfnamefont{F.~W.} \bibnamefont{Starr}},
  \bibinfo{author}{\bibfnamefont{F.}~\bibnamefont{Sciortino}},
  \bibinfo{author}{\bibfnamefont{S.~V.} \bibnamefont{Buldyrev}},
  \bibnamefont{and} \bibinfo{author}{\bibfnamefont{H.~E.}
  \bibnamefont{Stanley}}, \bibinfo{journal}{Phys. Rev. E}
  \textbf{\bibinfo{volume}{65}}, \bibinfo{pages}{041502}
  (\bibinfo{year}{2002}).

\bibitem[{\citenamefont{Louie et~al.}(1982)\citenamefont{Louie, Froyen, and
  Cohen}}]{pcec}
\bibinfo{author}{\bibfnamefont{S.~G.} \bibnamefont{Louie}},
  \bibinfo{author}{\bibfnamefont{S.}~\bibnamefont{Froyen}}, \bibnamefont{and}
  \bibinfo{author}{\bibfnamefont{M.~L.} \bibnamefont{Cohen}},
  \bibinfo{journal}{Phys. Rev. B} \textbf{\bibinfo{volume}{26}},
  \bibinfo{pages}{1738} (\bibinfo{year}{1982}).

\bibitem[{\citenamefont{Hamann}(1996)}]{Hamann}
\bibinfo{author}{\bibfnamefont{D.~R.} \bibnamefont{Hamann}},
  \bibinfo{journal}{Phys. Rev. Lett.} \textbf{\bibinfo{volume}{76}},
  \bibinfo{pages}{660} (\bibinfo{year}{1996}).

\end{thebibliography}

\end{document}